\RequirePackage{fix-cm}
\documentclass[smallcondensed]{svjour3}     % onecolumn (ditto)
\smartqed  % flush right qed marks, e.g. at end of proof
\pdfoutput=1
\usepackage{graphicx}
\usepackage{amsmath,amssymb}
\usepackage{url}
\begin{document}

\title{Cabello's nonlocality for generalized three-qubit \\ GHZ states%\thanks{Grants or other notes
}

\titlerunning{ Cabello's nonlocality for generalized three-qubit GHZ states}        % if too long for running head

\author{Jos\'{e} Luis Cereceda}


\institute{Jos\'{e} Luis Cereceda \at
              Telef\'{o}nica de Espa\~{n}a, Distrito Telef\'{o}nica, Edificio Este 1, 28050, Madrid, Spain \\             
              \email{jl.cereceda@movistar.es}           %  \\
}

\date{Received: date / Accepted: date}

\maketitle

\begin{abstract}
In this paper, we study Cabello's nonlocality argument (CNA) for three-qubit systems  configured in the generalized GHZ state. For
this class of states, we show that CNA runs for almost all entangled ones, and that the maximum probability of success of CNA is
14\% (approx.), which is attained for the maximally entangled GHZ state. This maximum probability is slightly higher than that
achieved for the standard Hardy's nonlocality argument (HNA) for three qubits, namely 12.5\%. Also, we show that the success
probability of both HNA and CNA for three-qubit systems can reach a maximum of $50\%$ in the framework of generalized no-signaling
theory.
\end{abstract}

\section{Introduction}
\label{sec:1}

Hardy's celebrated proof of nonlocality without inequalities for two-qubit systems caused much interest among physicists \cite{hardy1,hardy2}.
Hardy's proof has even been termed as ``the best version of Bell's theorem" \cite{mermin}. Soon after the publication of \cite{hardy1},
Pagonis and Clifton \cite{pagonis} extended Hardy's nonlocality argument (HNA) to $n$-qubit systems in a particular entangled state.
Moreover, Wu and Xie \cite{wu1} demonstrated Hardy's nonlocality for almost all entangled states of three qubits by using a particular type
of relationship among the coefficients of the given quantum state. Subsequently, Ghosh et al. \cite{ghosh} dispensed with such a restriction
and showed that HNA runs for all genuinely entangled states of three qubits. Furthermore, it was shown in \cite{ghosh} that for any maximally
entangled state of three qubits the probability of success of HNA can reach a maximum of 12.5\% (see also, in this respect, Refs.~\cite{wu2,cere1}
where this result is reproduced for the specific case of three qubits in the maximally entangled GHZ state). Actually, as shown in \cite{choud1},
this maximum turns out to be optimal over all pure entangled states of three qubits subjected to local projective measurements. This
result also holds for general (dichotomic) measurements on tripartite quantum systems of arbitrary dimension \cite{das}.

On the other hand, Cabello \cite{cabello} introduced a logical structure (of which Hardy's original formulation is a special case) to prove Bell's
theorem without inequalities for three-qubit GHZ and $W$ states. Cabello's logical structure is as follows (cf. Fig.~3 in \cite{cabello}):
Consider four events $A$, $B$, $C$, and $D$, of which $A$ and $C$ may happen in one system, and $B$ and $D$ may happen in another
(spatially separated) system. These events display the following features: (i) $A$ and $B$ sometimes happen, that is, the joint probability of
occurrence of $A$ and $B$ is nonzero; (ii) $A$ always implies $D$; (iii) $B$ always implies $C$; (iv) $C$ and $D$ happen with lower probability
than $A$ and $B$. Taken together, features (i)-(iv) are incompatible with local realism. The argument for nonlocality based on this logical structure
will be referred to as Cabello's nonlocality argument (CNA). Although CNA was originally intended for three-qubit systems, Liang and Li \cite{liang}
developed a variant of it to show nonlocality without inequalities for a class of two-qubit mixed states, and Kunkri and Choudhary \cite{kunkri1}
extended CNA to two spin-$s$ systems along the procedure followed in \cite{liang}. Moreover, Kunkri at al. \cite{kunkri2} showed that the
maximum probability of success of CNA for two-qubit systems in a pure entangled state is 0.1078, whereas, as is well known, the highest
probability of success of HNA for two-qubit systems is $(5\sqrt{5} - 11)/2 \approx 0.09$ \cite{hardy2}.

In this paper, we study CNA for three-qubit systems configured in the generalized GHZ-like state:
\begin{equation}\label{ghz}
|\psi \rangle = \frac{t}{\sqrt{1+t^2}}|v_1, v_2, v_3 \rangle + \frac{1}{\sqrt{1+t^2}}|w_1, w_2, w_3 \rangle ,
\end{equation}
where $\{ |v_k \rangle, |w_k \rangle \}$ is an arbitrary orthonormal basis in the state space of qubit $k$, $k = 1,2,3$. Without loss of generality,
we assume that $0 \leq t \leq 1$. Note that $t=0$ ($t=1$) corresponds to the product (maximally entangled) state. In Sect.~\ref{sec:2}, we show
that CNA runs for almost all entangles states of the form \eqref{ghz}. More precisely, it is shown that CNA does succeed for all the states \eqref{ghz}
for which $\epsilon < t \leq 1$, where $\epsilon$ is an infinitesimal quantity. The maximum probability of success $C_{\text{max}}$ of CNA
for the class of states \eqref{ghz} occurs at $t=1$, with $C_{\text{max}}$ being equal to $\frac{9}{64} = 0.140625$. In Sect.~\ref{sec:3}, we
discuss CNA for three-qubit systems in the framework of generalized no-signaling theory (GNST), that is, a theory that allows arbitrary correlations
between measurements on spatially separated systems as long as they are no-signaling \cite{popescu,masanes}. We find that the maximum success probability of both HNA and CNA within GNST reaches a value of $0.5$, in agreement with the result found in \cite{choud1} concerning HNA for
three-qubit systems in the context of GNST. Interestingly, this maximum value is the same as that attained for two-qubit systems under GNST \cite{choud1,cere2,ahanj,xiang}. We also give an explicit example of postquantum no-signaling correlations for three-qubit systems that satisfy the
modified Hardy-type nonlocality conditions established in \cite{rahaman}, and that violate maximally the {\it guess your neighbor's input} (GYNI)
inequality \cite{input1,input2}. We conclude with summary and some remarks in Sect.~\ref{sec:4}.

\section{Cabello's nonlocality argument for three qubits}
\label{sec:2}

Let us consider a $(3,2,2)$ Bell-type scenario in which three correlated qubits fly apart from a common source producing the triplet of qubits
in the state \eqref{ghz}. Each of the qubits then enters its own measuring station where, for each run of the experiment, one of two alternative measurements is performed: $U_1$ or $D_1$ for qubit 1, $U_2$ or $D_2$ for qubit 2, and $U_3$ or $D_3$ for qubit 3. Each measurement gives
the possible outcomes $+1$ or $-1$. The observables $U_k$ and $D_k$ have associated operators $\hat{U}_k =|u^{+}_{k}\rangle\langle
u^{+}_{k}| - |u^{-}_{k}\rangle\langle u^{-}_{k}|$ and  $\hat{D}_k =|d^{+}_{k}\rangle\langle d^{+}_{k}| - |d^{-}_{k}\rangle\langle
d^{-}_{k}|$, where the eigenvectors  $|u^{\pm}_k \rangle$ and $|d^{\pm}_k \rangle$ are related to the original basis vectors $|v_k \rangle$
and $|w_k \rangle$ by
\begin{align*}
& |u^{+}_{k}\rangle = \cos\alpha_k |v_k \rangle + e^{i\delta_k} \sin\alpha_k  |w_k \rangle,  \\
& |u^{-}_{k}\rangle = -e^{-i\delta_k}\sin\alpha_k |v_k \rangle + \cos\alpha_k  |w_k \rangle,\\
& |d^{+}_{k}\rangle = \cos\beta_k |v_k \rangle + e^{i\gamma_k} \sin\beta_k  |w_k \rangle,  \\
& |d^{-}_{k}\rangle = -e^{-i\gamma_k}\sin\beta_k |v_k \rangle + \cos\beta_k  |w_k \rangle .
\end{align*}
Consider next the following set of equations:
\begin{align}
& P(U_1,U_2,U_3|+++) = P,   \label{hc1} \\
& P(D_1,U_2,U_3|+++) = 0,  \label{hc2} \\
& P(U_1,D_2,U_3|+++) = 0,  \label{hc3} \\
& P(U_1,U_2,D_3|+++) = 0,  \label{hc4} \\
& P(D_1,D_2,D_3|---) = Q,     \label{hc5} 
\end{align}
where, for example, $P(U_1,U_2,U_3|+++)$ denotes the joint probability that a measurement of $U_1$, $U_2$, and $U_3$ on qubits 1, 2,
and 3, respectively, gives the outcome $+1$ for each of them. Eqs. \eqref{hc1}-\eqref{hc5} form the basis of Cabello's nonlocality
argument for three qubits. Indeed, it is easily seen that these equations contradict local realism whenever $Q < P$. For this, note first that
from Eqs. \eqref{hc2}-\eqref{hc4} we can deduce the following three statements: (1) if $D_1$, $U_2$, and $U_3$ are measured, then
necessarily $D_1 = -1$ if $U_2 = U_3 =+1$; (2) if $U_1$, $D_2$, and $U_3$ are measured, then necessarily $D_2 = -1$ if $U_1 = U_3 =
+1$; (3) if $U_1$, $U_2$, and $D_3$ are measured, then necessarily $D_3 = -1$ if $U_1 = U_2 = +1$. Now, from Eq.~\eqref{hc1}, we get
a fourth statement: (4) there is a nonzero probability $P$ of obtaining the results $U_1 = U_2 = U_3 =+1$ in a joint measurement of $U_1$,
$U_2$, and $U_3$. Then, combining the above four statements with the assumption of local realism, one is led to conclude that the
probability $P(D_1,D_2,D_3|---)$ should be at least $P$. But this contradicts Eq.~\eqref{hc5} if $Q < P$. The success probability for CNA can
therefore be quantified by the difference $P-Q $. In the particular case where $Q =0$ the above (Cabello) argument reduces to Hardy's.

For the state \eqref{ghz}, the quantum prediction for the joint probabilities \eqref{hc1} and \eqref{hc5} are given by
\begin{equation}\label{P}
P(t,\alpha_1,\alpha_2,\alpha_3,\delta) = \frac{t^2+ \tan^2 \alpha_1 \tan^2 \alpha_2 \tan^2 \alpha_3 +2t \cos\delta\tan\alpha_1 \tan
\alpha_2 \tan\alpha_3}{(1+t^2)(1+\tan^2\alpha_1)(1+\tan^2\alpha_2)(1+\tan^2\alpha_3)},
\end{equation}
and
\begin{equation}\label{Q}
Q(t,\beta_1,\beta_2,\beta_3,\gamma) = \frac{1+ t^2 \tan^2 \beta_1 \tan^2 \beta_2 \tan^2 \beta_3 -2t \cos\gamma\tan\beta_1
\tan\beta_2  \tan\beta_3}{(1+t^2)(1+\tan^2\beta_1)(1+\tan^2\beta_2)(1+\tan^2\beta_3)},
\end{equation}
respectively, where $\delta = \delta_1 + \delta_2 + \delta_3$ and $\gamma = \gamma_1 + \gamma_2 + \gamma_3$. On the other hand, in
order for the joint probabilities \eqref{hc2}, \eqref{hc3}, and \eqref{hc4} to vanish, it is necessary that
\begin{align}
& \gamma_1 + \delta_2 + \delta_3 = m_1 \pi,  \notag  \\
& \delta_1 + \gamma_2 + \delta_3 = m_2 \pi,  \notag  \\
& \delta_1 + \delta_2 + \gamma_3 = m_3 \pi,  \label{c1}
\end{align}
where $m_i = 0, \pm1, \pm2, \ldots\,$ ($i=1,2,3$). When the conditions in \eqref{c1} are met, the vanishing of the probabilities \eqref{hc2},
\eqref{hc3}, and \eqref{hc4} is equivalent to the fulfillment of the conditions
\begin{align}
& \tan\beta_1 \tan\alpha_2 \tan\alpha_3 = (-1)^{m_1 +1} t,   \notag  \\
& \tan\alpha_1 \tan\beta_2 \tan\alpha_3 = (-1)^{m_2 +1} t,   \notag  \\
& \tan\alpha_1 \tan\alpha_2 \tan\beta_3 = (-1)^{m_3 +1} t,   \label{c2}
\end{align}
respectively. Therefore, choosing $m_1 =m_2 =m_3 =+1$, from Eqs. \eqref{c1} it quickly follows that
\begin{equation}\label{r1}
\delta = \frac{1}{2} \big( 3\pi - \gamma \big),  
\end{equation}
Furthermore, from Eqs. \eqref{c2} it can be deduced that
\begin{align}
& \tan^2\alpha_1 = \frac{\tan\beta_1}{\tan\beta_2 \tan\beta_3}t,  \notag  \\
& \tan^2\alpha_2 = \frac{\tan\beta_2}{\tan\beta_1 \tan\beta_3}t,  \notag  \\
& \tan^2\alpha_3 = \frac{\tan\beta_3}{\tan\beta_1 \tan\beta_2}t , \label{r2}
\end{align}
where, without loss of generality, we assume that $0 \leq \alpha_i, \beta_i \leq \pi/2$. Thus, using relationships \eqref{r1} and \eqref{r2} in
Eq.~\eqref{P}, and employing the shorthand notation $ x \equiv \tan\beta_1$, $y \equiv \tan\beta_2$, and $z \equiv \tan\beta_3$, from
Eqs.~\eqref{P} and \eqref{Q} we finally obtain the success probability of CNA as
\begin{align}
C(t,x,y,z,\gamma) & = P(t,x,y,z,\gamma) - Q(t,x,y,z,\gamma)  \notag  \\
& =  \frac{(txyz)^2 \left[ t + xyz -2 \sqrt{txyz}\sin \left(\frac{\gamma}{2}\right)  \right]}
{(1+t^2)(xyz + tx^2)(xyz + ty^2)(xyz + tz^2)}  \notag \\
& - \frac{1 + (txyz)^2 - 2t xyz \cos\gamma}{(1+t^2)(1+x^2)(1+y^2)(1+z^2)}.  \label{cna}
\end{align}
A few comments are in order regarding the function $C(t,x,y,z,\gamma)$ in Eq. \eqref{cna}.
\begin{figure}[hhh]
\begin{center}
\vspace{-3.6cm}
\scalebox{0.43}{\includegraphics{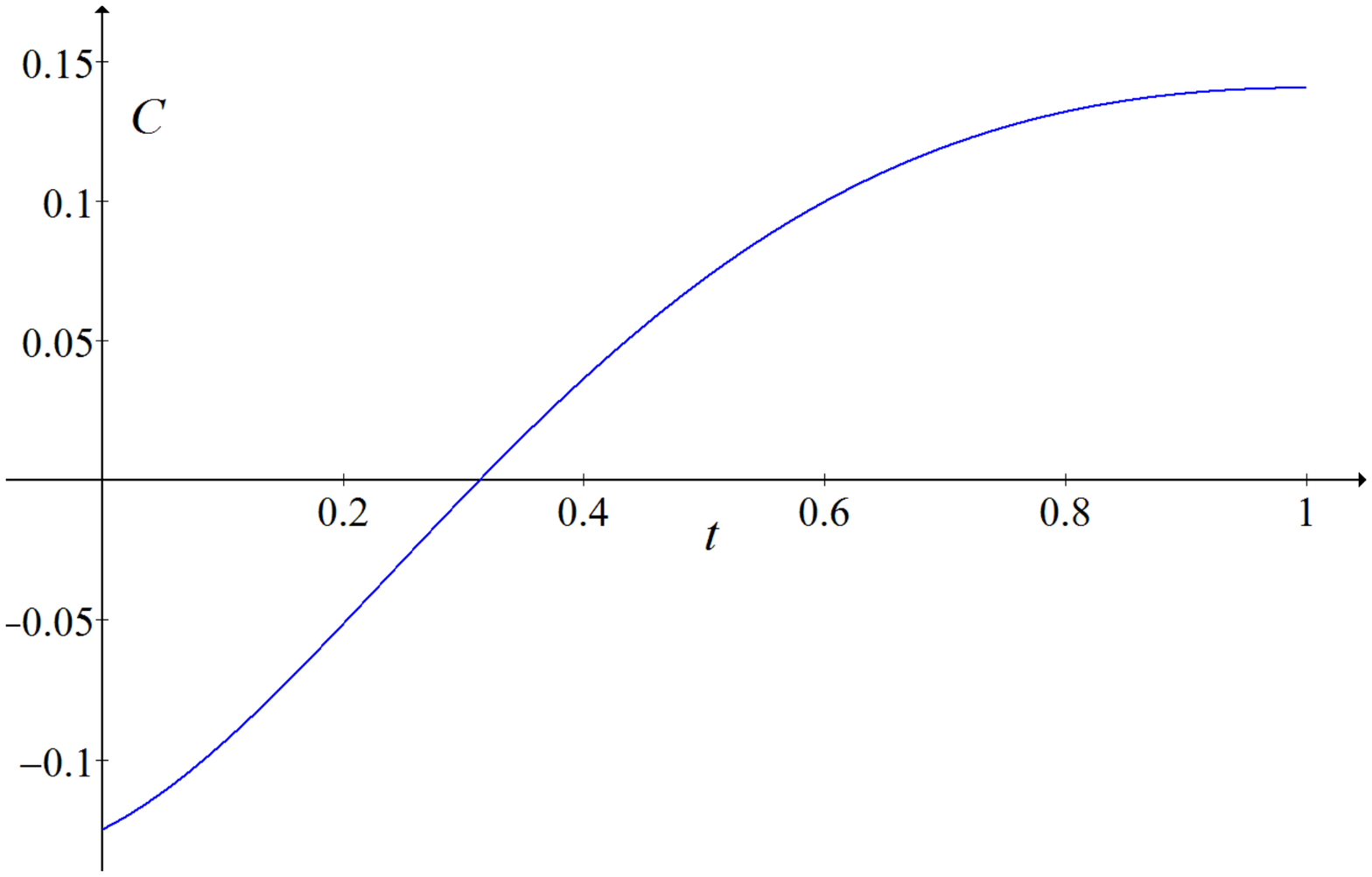}}
\vspace{-3.7cm}
\caption{\label{fig:1} Plot of $C(t,x,y,z,\gamma)$ for $x=y=z=1$ and $\gamma = -\arccos\frac{7}{8}$. The maximum $C_{\text{max}}=
0.140625$ is obtained for $t=1$.}
\vspace{-.5cm}
\end{center}
\end{figure}
\begin{enumerate}
\item $C(t,x,y,z,\gamma)$ remains invariant under the permutations of the elements of the set $\{ x,y,z \}$. This means that in order for
$C(t,x,y,z,\gamma)$ to attain an extremum value it is necessary that $x=y=z$. Indeed, it can be checked numerically that, for $0 \leq t \leq 1$ and
$x,y,z \geq 0$, the global maximum of $C(t,x,y,z,\gamma)$ is obtained for $t=1$, $x=y=z=1$, and $\gamma = \gamma_0 = - \arccos\frac{7}{8}
\approx -28.955^{\circ}$, and is given by $C_{\text{max}} = P_{\text{max}} - Q_{\text{min}}= \frac{10}{64} - \frac{1}{64} = \frac{9}{64} =
0.140625$. In Fig.~\ref{fig:1}, we have plotted $C(t,1,1,1,\gamma_0)$ as a function of $t$. Note that $C(t,1,1,1,\gamma_0) <0$ for $0 \leq t
\lessapprox 0.3126$, and thus, for $x=y=z=1$ and $\gamma =\gamma_0$, CNA does not work for $t$ ranging in the above interval.

\item For $t=0$, we have
\begin{equation*}
C(0,x,y,z,\gamma) = -\frac{1}{(1+x^2)(1+y^2)(1+z^2)}  < 0,
\end{equation*}
so that CNA is not applicable for the product state irrespective of the values of $x,y,z,$ and $\gamma$. Due to continuity of $C(t,x,y,z,\gamma)$,
there is inevitably an interval $0 \leq t < \epsilon$ for which $C(t,x,y,z,\gamma) <0$. The parameter $\epsilon$ can be made as small as desired
by taking the product $xyz$ large enough. It should be noted, however, that, for large values of $xyz$, $C(t,x,y,z,\gamma)$ is nonnegative only for
a negligibly small interval of $t$. A better option is to take the product $xy$ large enough while keeping $z$ relatively small. In Fig.~\ref{fig:2}, we
have plotted $C(t,10^4,10^4,0.2,\gamma_0)$ as a function of $t$. For this case, CNA runs for all $t$ in the interval $10^{-8} \leq t \lessapprox
0.8198$. As another example, let us mention that $C(t,10^2,10^2,0.02,\gamma_0)$ is positive for $10^{-4} \leq t \leq 1$. The maximum obtained
in this case is, however, not more than $0.00031$.
\begin{figure}[bbb]
\begin{center}
\vspace{-4cm}
\scalebox{0.43}{\includegraphics{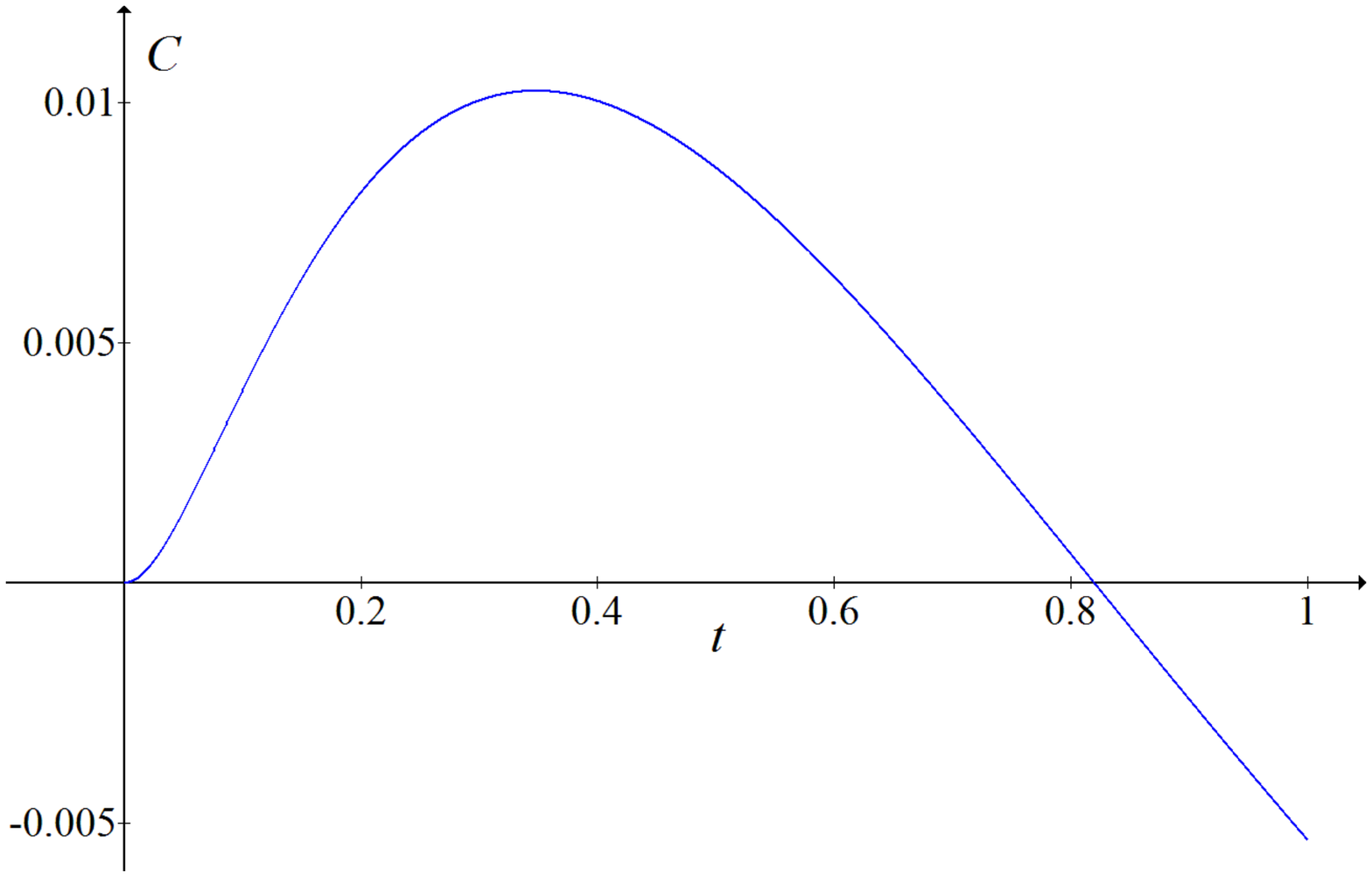}}
\vspace{-3.7cm}
\caption{\label{fig:2} Plot of $C(t,x,y,z,\gamma)$ for $x=y=10^4$, $z=0.2$, and $\gamma = -\arccos\frac{7}{8}$. For $t=0$, the displayed
function attains a negative value of $\approx -10^{-16}$.}
\vspace{-.6cm}
\end{center}
\end{figure}

\item For the special case in which $txyz =1$ and $\gamma = 0^{\circ}$, we have $Q(t,x,y,z,\gamma)=0$ and then CNA reduces to HNA. Taking $x=y=z=t^{-1/3}$, we obtain
\begin{equation*}
C(t, t^{-1/3}, t^{-1/3}, t^{-1/3}, 0^{\circ}) = \frac{t^2}{ \left( 1 + t^{4/3} \right)^3},
\end{equation*}
which attains a maximum value of $\frac{1}{8} = 0.125$ for $t=1$, in accordance with the results in \cite{ghosh,wu2,cere1}.

\end{enumerate}

It is worth noting that, out of the five probabilities appearing in Eqs. \eqref{hc1}-\eqref{hc5}, one can derive the following Bell-type inequality
\cite{cere1,choud2,wang,yu,guney,home}
\begin{align}
P(U_1,U_2,U_3|+++) - P(D_1,D_2,D_3|---)  & \leq  P(D_1,U_2,U_3|+++)  \notag  \\
& + P(U_1,D_2,U_3|+++)  \notag  \\
& + P(U_1,U_2,D_3|+++).  \label{ineq1}
\end{align}
When the constraints in Eqs. \eqref{hc2}-\eqref{hc4} are satisfied, inequality \eqref{ineq1} reduces to
\begin{equation}\label{ineq2}
C = P - Q = P(U_1,U_2,U_3|+++) - P(D_1,D_2,D_3|---) \leq 0,
\end{equation}
which is violated whenever $C >0$. Furthermore, for the case in which $Q=0$, the maximum quantum violation $P=\frac{1}{8}$ of inequality
\eqref{ineq2} is obtained for a state that is local-unitarily connected to the three-qubit GHZ state \cite{choud1}. Motivated by this fact, we
conjecture that, for the case in which $Q >0$, the maximum quantum violation of inequality \eqref{ineq2} (subject to the fulfillment of conditions \eqref{hc2}-\eqref{hc4}) is achieved, over all possible states and choice of observables, for the GHZ state $|\psi \rangle =\frac{1} {\sqrt{2}}(|v_1,
v_2,v_3 \rangle + |w_1, w_2, w_3 \rangle)$, with the maximum violation being equal to $C_{\text{max}} =\frac{9}{64}$.

\section{Cabello's nonlocality argument for three qubits in generalized no-signaling theory}
\label{sec:3}

We now study the optimal success probability of CNA for three-qubit systems within the framework of generalized no-signaling theory (GNST)
\cite{popescu,masanes}. This is a minimally constrained theory for which, in addition to the consistency conditions of positivity and normalization
of the probability distributions, we assume the no-signaling condition which forbids faster-than-light communication between distant observers. Let us
first recall that, for two-qubit systems, the success probability of both HNA and CNA can be increased up to $0.5$ within GNST \cite{choud1,cere2,ahanj,xiang}. For two-qubit systems, the Cabello-type nonlocality conditions can be written as \cite{kunkri2}
\begin{equation}\label{cabcon}
\begin{aligned}
& P(U_1,U_2|++) = R,   \\
& P(D_1,U_2|++) = 0,   \\
& P(U_1,D_2|++) = 0,   \\
& P(D_1,D_2|--) = S, 
\end{aligned}
\end{equation}
with $C_2 = R -S >0$. By using the above-mentioned conditions of normalization and no-signaling, and taking into account the constraints
$P(D_1,U_2|++) = P(U_1,D_2|++) = 0$, it is easily shown that
\begin{equation}\label{chsh}
\text{CHSH} = 2 +4 C_2,
\end{equation}
where $\text{CHSH} = E(U_1, U_2) - E(D_1, U_2) - E(U_1, D_2) - E(D_1, D_2)$ is the Clauser-Horne-Shimony-Holt sum of correlations \cite{clauser},
and where, for example, $E(U_1,U_2) = P(U_1,U_2|++) + P(U_1, U_2|--) - P(U_1, U_2|+-) - P(U_1, U_2|-+)$. Since, by definition, $\text{CHSH}$
cannot exceed the algebraic limit of $4$, it is concluded that $C_2$ can reach a maximum value of $0.5$ without violating the no-signaling constraint. Furthermore, since a value of $\text{CHSH} =4$ necessarily implies that $E(U_1, U_2)=+1$ and $E(D_1, D_2) =-1$, this maximum is obtained for $R
= 0.5$ and $S =0$. Incidentally, we also note that, as is apparent from Eq.~\eqref{chsh}, the Bell-CHSH inequality, $\text{CHSH} \leq 2$ \cite{clauser},
is violated whenever $C_2 >0$ \cite{xiang}.

Let us now consider the set of $64$ joint probabilities for three-qubit systems\linebreak $P(X_1,Y_2,Z_3|ijk)$, where $(X_1,Y_2,Z_3) \in \{ U_1,D_1 \}
\times \{ U_2,D_2 \} \times \{ U_3,D_3 \}$ and $i,j,k \in \{+,- \}$. In the framework of a general probabilistic theory, the allowed probability
distributions $\{ P(X_1,Y_2,Z_3|ijk) \}$ are required to satisfy the following conditions \cite{popescu,masanes}:
\begin{itemize}
\item Positivity:
\begin{equation}\label{pos}
P(X_1,Y_2,Z_3|ijk) \geq 0, \,\,\, \forall\,  X_1,Y_2,Z_3,i,j,k.
\end{equation}
\item Normalization:
\begin{equation}
\sum_{i,j,k} P(X_1,Y_2,Z_3|ijk) =1, \,\,\, \forall\,    X_1,Y_2,Z_3.
\end{equation}
\item No-signaling:
\begin{align}
& \sum_i  P(U_1,Y_2,Z_3|ijk) = \sum_i  P(D_1,Y_2,Z_3|ijk),  \,\,\, \forall\,  Y_2,Z_3,j,k,  \notag \\
& \sum_j  P(X_1,U_2,Z_3|ijk) = \sum_j  P(X_1,D_2,Z_3|ijk),  \,\,\, \forall\,  X_1,Z_3,i,k,  \notag \\
& \sum_k  P(X_1,Y_2,U_3|ijk) = \sum_k  P(X_1,Y_2,D_3|ijk),  \,\,\, \forall\,  X_1,Y_2,i,j.
\end{align}
\end{itemize}
In addition to this, we impose the constraints in Eqs. \eqref{hc1}-\eqref{hc5}, namely
\begin{itemize}
\item Hardy/Cabello nonlocality conditions:
\begin{equation}\label{hcab1}
C = P - Q = P(U_1,U_2,U_3|+++) - P(D_1,D_2,D_3|---) > 0,
\end{equation}
and
\begin{equation}\label{hcab2}
P(D_1,U_2,U_3|+++) = P(U_1,D_2,U_3|+++) = P(U_1,U_2,D_3|+++) = 0.
\end{equation}
\end{itemize}
Our goal is to maximize the difference $C = P-Q$ subject to the fulfillment of all the conditions in Eqs. \eqref{pos}-\eqref{hcab2}. The software
program Mathematica gives us the result $C =0.5$, along with the particular solution set
\begin{equation}\label{set1}
\begin{aligned}
& P(U_1,U_2,U_3|+++) = P(U_1,U_2,U_3|-+-) = 0.5,    \\
& P(U_1,U_2,D_3|++-) = P(U_1,U_2,D_3|-++) = 0.5,    \\
& P(U_1,D_2,U_3|+-+) = P(U_1,D_2,U_3|---) = 0.5,      \\
& P(U_1,D_2,D_3|+--) = P(U_1,D_2,D_3|--+) = 0.5,     \\
& P(D_1,U_2,U_3|++-) = P(D_1,U_2,U_3|-++) = 0.5,    \\
& P(D_1,U_2,D_3|++-) = P(D_1,U_2,D_3|-++) = 0.5,    \\
& P(D_1,D_2,U_3|+--) = P(D_1,D_2,U_3|--+) = 0.5,    \\
& P(D_1,D_2,D_3|+--) = P(D_1,D_2,D_3|--+) = 0.5, 
\end{aligned}
\end{equation}
with the remaining $48$ joint probabilities being all zero. The set of probabilities \eqref{set1} was already obtained by Choudhary et al. in
\cite{choud1}, where it is further noted that such solution is not unique. Another set of probabilities yielding $C =0.5$ and which satisfies all the
conditions in Eqs. \eqref{pos}-\eqref{hcab2} is, for example,
\begin{equation}\label{set2}
\begin{aligned}
& P(U_1,U_2,U_3|+++) = P(U_1,U_2,U_3|--+) = 0.5,   \\
& P(U_1,U_2,D_3|++-) = P(U_1,U_2,D_3|---) = 0.5,    \\
& P(U_1,D_2,U_3|+-+) = P(U_1,D_2,U_3|-++) = 0.5,   \\
& P(U_1,D_2,D_3|+--) = P(U_1,D_2,D_3|-+-) = 0.5,    \\
& P(D_1,U_2,U_3|+-+) = P(D_1,U_2,U_3|-++) = 0.5,    \\
& P(D_1,U_2,D_3|+--) = P(D_1,U_2,D_3|-+-) = 0.5,    \\
& P(D_1,D_2,U_3|+-+) = P(D_1,D_2,U_3|-++) = 0.5,  \\
& P(D_1,D_2,D_3|+--) = P(D_1,D_2,D_3|-+-) = 0.5,  
\end{aligned}
\end{equation}
where only the nonzero probabilities have been written out. It is important to notice that the system of linear equations \eqref{pos}-\eqref{hcab2}
is incompatible if we assume that $P$ and $Q$ are of the form $P = 0.5 + \delta$ and $Q = \delta$, with $\delta >0$. The maximum value $C=
0.5$ allowed by GNST should therefore be realized for $P =0.5$ and $Q=0$. Thus, regarding the maximum of $C$, CNA does not entail any advantage 
over HNA in the context of GNST. This contrasts with the theory of quantum mechanics, which, as we have seen, predicts a maximum value of $C$ for
CNA which is greater than that predicted for HNA.

It is worth pointing out, on the other hand, that probability distributions like those in Eqs. \eqref{set1} and \eqref{set2} satisfy any one of the
Svetlichny-type inequalities \cite{svet,cere3,mitchell,bancal} such as, for example,
\begin{align*}
S_v & = | E(U_1,U_2,U_3) + E(D_1,U_2,U_3) + E(U_1,D_2,U_3) + E(U_1,U_2,D_3)  \\
& - E(U_1,D_2,D_3) - E(D_1,U_2,D_3) - E(D_1,D_2,U_3) - E(D_1,D_2,D_3)| \leq 4,
\end{align*}
where $E(X_1,Y_2,Z_3) = \sum_{i,j,k} ijk P(X_1,Y_2,Z_3|ijk)$ is the expectation value of the product of the measurement outcomes of the
observables $X_1$, $Y_2$, and $Z_3$. Indeed, both probability distributions \eqref{set1} and \eqref{set2} pertain to the only class of extremal
two-way local correlations that are not fully local \cite{popescu}, which means that such correlations can be accounted for by a hybrid hidden
variables model in which arbitrary (though no-signaling) correlations may take place between two of the qubits, but only local correlations are present between these two qubits and the third one \cite{svet,cere3,mitchell,bancal}. Specifically, for the probability distribution \eqref{set2}, it is readily seen
that, when qubit 3 is ignored (traced out), the correlations between qubits 1 and 2 are maximally nonlocal (in fact, they make the parameter CHSH as
large as $4$), while the correlations between the subsystem of qubits 1 and 2, on the one hand, and qubit 3, on the other hand, are strictly local. Note,
however, that both probability distributions \eqref{set1} and \eqref{set2} violate maximally the standard Bell-type inequality \eqref{ineq1} when the
conditions in Eq.~\eqref{hcab2} are met.

The existence of multiple solutions to Eqs. \eqref{pos}-\eqref{hcab2} yielding $C=0.5$, stems from the fact that, for the case where $P=0.5$ and
$Q=0$, the set of nonlocality conditions \eqref{hc1}-\eqref{hc5} (along with the requirements of positivity, normalization, and no-signaling) only
determines five out of the eight correlations $E(X_1,Y_2,Z_3)$. This contrasts with the two-qubit case where the nonlocality conditions \eqref{cabcon}
fix all four correlations $E(X_1,Y_2)$ so that the probability distribution for two-qubit systems yielding $C_2 =0.5$ is (for the said conditions
\eqref{cabcon}) unique.

Next we give the following probability distribution fulfilling all the conditions in Eqs.~\eqref{pos}-\eqref{hcab2}, and for which $C = \frac{1}{3}$
(with $P= \frac{1}{3}$ and $Q=0$):
\begin{equation}\label{set3}
\begin{aligned}
& P(U_1,U_2,U_3|+++) = P(U_1,U_2,U_3|-+-) = P(U_1,U_2,U_3|--+) = 1/3,   \\
& P(U_1,U_2,D_3|++-) = P(U_1,U_2,D_3|-++) = P(U_1,U_2,D_3|---) =1/3,    \\
& P(U_1,D_2,U_3|+-+) = P(U_1,D_2,U_3|-+-) = P(U_1,D_2,U_3|--+) =1/3,   \\
& P(U_1,D_2,D_3|+--) = P(U_1,D_2,D_3|-+-) =  P(U_1,D_2,D_3|--+) =1/3,    \\
& P(D_1,U_2,U_3|+-+) = P(D_1,U_2,U_3|-++) = P(D_1,U_2,U_3|-+-) = 1/3,    \\
& P(D_1,U_2,D_3|+--) = P(D_1,U_2,D_3|-++) = P(D_1,U_2,D_3|-+-) = 1/3,    \\
& P(D_1,D_2,U_3|+-+) = P(D_1,D_2,U_3|-+-) = P(D_1,D_2,U_3|--+) = 1/3,  \\
& P(D_1,D_2,D_3|+--) = P(D_1,D_2,D_3|-+-) =  P(D_1,D_2,D_3|--+) = 1/3,  
\end{aligned}
\end{equation}
with the remaining $40$ joint probabilities being all zero. In addition to the nonlocality conditions \eqref{hcab1}-\eqref{hcab2}, the set of probabilities \eqref{set3} satisfies the modified Hardy-type nonlocality conditions established in \cite{rahaman}, which, for the specific case of three-qubit systems,
read as
\begin{equation}\label{rah}
\begin{aligned}
P(U_1,U_2,U_3 & |+++) >0,  \\
P(D_1,U_2 & |++) =0,  \\
P(D_2,U_3 & |++) =0,  \\
P(U_1,D_3 & |++) =0,  \\
P(D_1,D_2,D_3 & |---) =0,
\end{aligned}
\end{equation}
where $P(D_1,U_2|++)$, $P(D_2,U_3|++)$, and $P(U_1,D_3|++)$ denote marginal probabilities. The value $C = \frac{1}{3}$ is the maximum
of $C$ achievable within GNST and which is consistent with the conditions in Eqs.~\eqref{rah}. Moreover, as shown in \cite{rahaman}, only a unique
pure genuinely entangled three-qubit state satisfies the above conditions \eqref{rah}. It is therefore concluded that, in contrast with the probability
distributions \eqref{set1} and \eqref{set2}, the one in \eqref{set3} must violate some Svetlichny-type inequality, namely
\begin{align*}
S_v & = | E(U_1,U_2,U_3) - E(U_1,U_2,D_3) + E(U_1,D_2,U_3) + E(U_1,D_2,D_3)  \\
& - E(D_1,U_2,U_3) + E(D_1,U_2,D_3) + E(D_1,D_2,U_3) + E(D_1,D_2,D_3)| \leq 4,
\end{align*}
with $S_v = \frac{16}{3} \approx 5.333$. Let us further observe that the set of probabilities \eqref{set3} violates maximally the GYNI inequality
\cite{input1,input2}
\begin{align}
P(U_1,U_2,U_3|+++) &+ P(U_1,D_2,D_3|--+)   \notag \\
& + P(D_1,U_2,D_3|+--) + P(D_1,D_2,U_3|-+-) \leq 1. \label{gyni}
\end{align}
Indeed, as shown in \cite{input1,input2}, inequality \eqref{gyni} is obeyed by quantum theory, while it is maximally violated by those postquantum
no-signaling correlations for which each one of the four probabilities in \eqref{gyni} is equal to $\frac{1}{3}$. Note, on the other hand, that the (postquantum) probability distributions \eqref{set1} and \eqref{set2} satisfy the above GYNI inequality.

To end this section, we provide yet another probability distribution fulfilling all the conditions Eqs.~\eqref{pos}-\eqref{hcab2}, for which $C = 0.4$
(with $P= 0.6$ and $Q=0.2$):
\begin{align*}
& P(U_1,U_2,U_3|+++) = 0.6,  \quad P(U_1,U_2,U_3|--+) = 0.4,   \\
& P(U_1,U_2,D_3|++-) = 0.6,   \quad  P(U_1,U_2,D_3|---) = 0.4,    \\
& P(U_1,D_2,U_3|+-+) = 0.6,   \quad  P(U_1,D_2,U_3|-++) = 0.4,   \\
& P(U_1,D_2,D_3|+--) = 0.6,   \quad  P(U_1,D_2,D_3|-+-) = 0.4,    \\
& P(D_1,U_2,U_3|+-+) = 0.4,  \quad  P(D_1,U_2,U_3|-++) = 0.6,    \\
& P(D_1,U_2,D_3|+--) = 0.4,  \quad P(D_1,U_2,D_3|-+-) = 0.6,    \\
& P(D_1,D_2,U_3|+-+) = 0.4,  \quad P(D_1,D_2,U_3|-++) = 0.4  \\
& P(D_1,D_2,U_3|--+) = 0.2,  \quad P(D_1,D_2,D_3|+--) = 0.4  \\
& P(D_1,D_2,D_3|-+-) = 0.4, \quad  P(D_1,D_2,D_3|---) = 0.2,  
\end{align*}
with the remaining 46 joint probabilities being all zero.

\section{Conclusions}
\label{sec:4}

In this paper, we have fully characterized Cabello's nonlocality for three qubits in the generalized GHZ-like state \eqref{ghz}. For this class of states
the maximum success probability of CNA is found to be $\frac{9}{64}$, which is achieved for the maximally entangled GHZ state ($t=1$). We
conjectured that the value $\frac{9}{64}$ is indeed optimal over all entangled states of three qubits subjected to arbitrary local projective
measurements. Moreover, we have shown that CNA runs for almost all entangled states of the form \eqref{ghz}. Specifically, for all $t$ in the interval
$\epsilon < t \leq 1$ (where $\epsilon$ is an infinitesimal quantity), there exist observables $U_k$ and $D_k$ ($k =1,2,3$) for which the conditions in
Eqs. \eqref{hc1}-\eqref{hc5} are satisfied, with $C = P-Q >0$. Moreover, we have shown that, for three-qubit systems, the maximum success probability
of both HNA and CNA can go up to $0.5$ within GNST. In this respect, we have pointed out that all the probability distributions giving $C = P -Q =0.5$
under GNST satisfy $P= 0.5$ and $Q =0$. We have also given an explicit (postquantum) probability distribution which violates maximally the GYNI
inequality \eqref{gyni}, and which satisfies the stronger nonlocality conditions in Eqs.~\eqref{rah} (as compared to the weaker nonlocality conditions \eqref{hcab2} involved in either HNA or CNA).

On the other hand, it was shown in \cite{cere1} that for $n$-qubit systems ($n \geq 3$) in the generalized GHZ-like state
\begin{equation*}
|\psi \rangle = \frac{t}{\sqrt{1+t^2}}|v_1, v_2, \ldots, v_n \rangle + \frac{1}{\sqrt{1+t^2}}|w_1, w_2,\ldots, w_n \rangle, 
\end{equation*}
the maximum success probability $P_{n}^{\text{max}}$ of HNA is obtained for $t =1$, and is given by
\begin{equation}\label{exp}
P_{n}^{\text{max}}(n \geq 3) = \left( \frac{1}{2^n}\right) \left[ 1+ \cos\left(\frac{\pi}{n-1}\right)\right].
\end{equation}
From Eq. \eqref{exp}, we can see that $P_{n}^{\text{max}}$ decreases exponentially with $n$, and thus it seems not very promising to investigate
Cabello-type nonlocality for $n$-qubit systems with $n\geq 4$. Regarding the ladder version of HNA involving two-qubit systems and $K+1$ observables
per qubit \cite{hardy3,hardy4}, it can be shown that, interestingly, the maximum success probability of CNA (call it $C_{K}^{\text{max}}$) for this
scenario is strictly greater than that corresponding to HNA (call it $P_{K}^{\text{max}}$) for all $K=1,2,\ldots\,$ \cite{cere4}. Table \ref{tb:1} lists $P_{K}^{\text{max}}$ and $C_{K}^{\text{max}}$ for $K=1$ to $10$. Furthermore, as $K \to\infty$, both $P_{K}^{\text{max}}$ and $C_{K}^{
\text{max}}$ tend to $0.5$, which is the maximum allowed value within GNST for any given $K$ \cite{cere5,cere6}. Finally, it is to be mentioned that
Chen et al. \cite{kwek} found a generalization of HNA applicable to high dimensional bipartite quantum systems. It seems to be worthwhile studying
CNA for this kind of scenario, in particular for the relatively simple case of two-qutrit systems.

\renewcommand{\arraystretch}{1.2}
\begin{table}[ttt]
\centering
\vspace{.6mm}
\begin{tabular}{ccc|ccc}\hline
$K$ & $P_{K}^{\text{max}}$ &  $C_{K}^{\text{max}}$ & $K$  &  $P_{K}^{\text{max}}$ &  $C_{K}^{\text{max}}$ \\ \hline\hline
$K=1$ & 0.090169 & 0.107813 & $K =6$ & 0.322175 & 0.324612   \\
$K=2$ & 0.174550 & 0.185185 & $K =7$ & 0.339721 & 0.341612   \\
$K=3$ & 0.231263 & 0.237964 & $K =8$ & 0.353936 & 0.355444   \\
$K=4$ & 0.270880 & 0.275415 & $K =9$ & 0.365697 & 0.366926   \\
$K=5$ & 0.299953 & 0.303204 & $K =10$ & 0.375597 & 0.376617   \\  \hline
\end{tabular}
\vspace{1.5mm}
\caption{\label{tb:1} Numerical values of $P_{K}^{\text{max}}$ and $C_{K}^{\text{max}}$ for $K =1$ to $10$ (see text). The value $C_{1}^{
\text{max}} = 0.1078$ was derived in Ref.~\cite{kunkri2}.}
\vspace{-0.4cm}
\end{table}

%\begin{figure}
% \includegraphics{example.eps}
%\end{figure}
%\begin{figure*}
%  \includegraphics[width=0.75\textwidth]{example.eps}
%\end{figure*}
%



\end{document}